\documentstyle[12pt]{article}
\catcode`\@=11

\@addtoreset{equation}{section}
\def\ksection{\arabic{section}}

\def\@normalsize{\@setsize\normalsize{15pt}\xiipt\@xiipt
           \abovedisplayskip 14pt plus3pt minus3pt%
           \belowdisplayskip \abovedisplayskip
           \abovedisplayshortskip  \z@ plus3pt%
           \belowdisplayshortskip  7pt plus3.5pt minus0pt}

\def\small{\@setsize\small{13.6pt}\xipt\@xipt
           \abovedisplayskip 16pt plus3pt minus3pt%
           \belowdisplayskip \abovedisplayskip
           \abovedisplayshortskip  \z@ plus3pt%
           \belowdisplayshortskip  7pt plus3.5pt minus0pt
             \def\@listi{\parsep 4.5pt plus 2pt minus 1pt
             \itemsep \parsep
             \topsep 9pt plus 3pt minus 3pt}}

\def\underline#1{\relax\ifmmode\@@underline#1\else
        $\@@underline{\hbox{#1}}$\relax\fi}
\@twosidetrue

\def\thesection{\Roman{section}.}

\long\def\@makecaption#1#2{
 \vskip 10pt
 \setbox\@tempboxa\hbox{#1: #2}
 \ifdim \wd\@tempboxa >\hsize #1: #2\par \else \hbox
        to\hsize{\box\@tempboxa\hfil}
 \fi}

\catcode`\@=12

\relax

\newcounter{appendix}
\def\appendix{\par
 \addtocounter{appendix}{1}
 \def\thesection{Appendix \Alph{appendix}:}
 \def\ksection{\Alph{appendix}}}

\newskip\humongous \humongous=0pt plus 1000pt minus 1000pt

\newif\ifdtup

\def\oldreffmt#1{\rlap{[#1]} \hbox to 2\parindent{}}

\def\figfmt#1{\rlap{Figure {#1}} \hbox to 1in{}}

\def\etal{\hbox{\it et al.}}

\def\slash#1{#1\!\!\!/\!\,\,}
\def\beq{\begin{equation}}
\def\eeq{\end{equation}}
\def\bea{\begin{eqnarray}}
\def\eea{\end{eqnarray}}

\def\bq{\begin{quote}}
\def\eq{\end{quote}}

\def \lta {\mathrel{\vcenter
     {\hbox{$<$}\nointerlineskip\hbox{$\sim$}}}}
\def \gta {\mathrel{\vcenter
     {\hbox{$>$}\nointerlineskip\hbox{$\sim$}}}}
\def \etal {{\it et al.}\ }
\relax

\def\SM{Standard Model }

\def\SMp{Standard Model. }

\def\EW{electro--weak }

\def\GB{Goldstone Boson }

\def\GBs{Goldstone Bosons }

\def\DSB{Dynamical Symmetry Breaking }

\def\bar{\overline}
\def\phi{\varphi}

\def\slimits{\mbox{\tiny
        $\begin{array}{r@{\hspace{0.mm}}c@{\hspace{0.mm}}l}
                \Sigma_1 & = & \Sigma_t \\
                \Sigma_2 & = & 0
        \end{array}$}}

\def\addcontentsline#1#2#3{}

\begin{document}

%    the titlepage

{    % start of local definitions
\def\thefootnote{\fnsymbol{footnote}}
\thispagestyle{empty}

\ \vskip -.8cm
\ \hskip 12.1cm HD--THEP--92--55

\ \vskip -1.2cm

\vskip 2.3cm
\begin{center}
      {\Large\sc\bf Custodial $SU(2)$ Violation and the Origin of}\\
      \ \\
      {\Large\sc\bf Fermion Masses}\\
\vskip 1.7cm
      {A. Blumhofer\ \  and\ \  M. Lindner\footnote{Heisenberg Fellow}}\\

\vskip .8cm
      {\sl  Institut f\"ur Theoretische Physik\\
      der Universit\"at Heidelberg\\
      Philosophenweg 16, D--W--6900 Heidelberg}\\
\end{center}

% local footnote without symbol
{\def\thefootnote{}
 \footnote{Email: T36 (A.B.) or Y29 (M.L.) at VM.URZ.UNI-HEIDELBERG.DE}
}

\vskip 1.6cm
   \begin{center}{\Large\bf Abstract}\end{center}
\par \vskip .05in
Custodial $SU(2)$ breaking due to dynamical fermion masses is studied
in a rather general context and it is shown how some well known limiting
cases are correctly described. The type of ``gap equation'' which can
systematically lead to extra negative contributions to the so--called
$\rho$--parameter is emphasized. Furthermore general model independent
features are discussed and it is shown how \EW precision measurements
can be sensitive to the fermion content and/or dynamical features of
a given theory.

} % end all local definitions

\newpage
%------- body of paper ---------------------------------------------

% reset counters
\setcounter{page}{1}
\setcounter{footnote}{0}

\section{Introduction}

The \SM of \EW interactions is today in very good shape even though the
Higgs mechanism is for a number of reasons unsatisfactory. The model agrees
however with all known experimental facts and there is even evidence for
quantum corrections. On the other hand a Higgs particle has not yet been
found and the symmetry breaking mechanism is untested. Besides the vacuum
expectation value (given by the Fermi constant) other essential experimental
information is expressed by model independent parametrizations of radiative
corrections in terms of the so--called $S$, $T$, $U$ variables \cite{PesTa},
where $T$ is related to the old $\rho$--parameter \cite{rhoinvent} by
$\alpha (T-T_0)=\rho-1=\Delta\rho$ (where $\alpha=e^2/4\pi$). This
$\rho$--parameter (which is experimentally very close to unity) is actually
defined in terms of the charged and neutral \EW \GB decay constants
$F_\pm(0)$ and $F_3(0)$ as
\beq
\rho:=\frac{F_\pm^2(0)}{F_3^2(0)} = 1 + \Delta\rho~;\quad
0 \lta \Delta\rho \lta 0.01~,
\label{defrho}
\eeq
and $T\equiv T_0$, i.e. $\rho\equiv 1$, can be understood in terms of an
extra global ``custodial'' symmetry transforming charged and neutral \GBs
into each other such that $F_\pm$ and $F_3$ must be identical. Small
deviations from $\rho=1$ are perturbations of this symmetry and this article
deals with such deviations due to a dynamical origin of fermion masses.

In the \SM the four real components of the Higgs doublet $\Phi$ correspond
to a global $SO(4) \simeq SU(2)_L\times SU(2)_R$ invariance of the pure
scalar Lagrangian with an extra custodial $SU(2)$ symmetry. This can be
made explicit by defining the matrix field $\Omega:=(\tilde{\Phi},\Phi)$
which transforms as $\Omega \rightarrow U_L\Omega U_R^+$, where
$\tilde{\Phi}=-i\sigma_2\Phi^*$ and $U_{L/R}:=exp(i\tau_a\lambda_a^{L/R})$.
Due to $\Phi^+\Phi = 1/2~Tr(\Omega^+\Omega)$ those parts
of the Lagrangian which depend only on $\Phi^+\Phi$ possess an extra
$SU(2)$ symmetry. If this were an exact symmetry of the full Lagrangian
then it would guarantee exactly (i.e. to all orders) $\rho\equiv 1$. The \SM
contains however two sources of custodial $SU(2)$ violations outside of
the pure Higgs sector, namely the $U(1)$ hypercharges and the asymmetries
of Yukawa couplings. In terms of $\Omega$ these custodial $SU(2)$
violating pieces can be written as
\beq
\delta{\cal L}_{custodial} = - g_1B_\mu Re
\left\{Tr\left[\tau_3\Omega^+(D_2^\mu\Omega)\right]\right\}
- \left(\frac{g_t-g_b}{2}\right)
\left(\bar{L}\Omega\tau_3R+\bar{R}\tau_3\Omega^+L\right)~,
\label{deltaL}
\eeq
where
$D_2^\mu =\partial^\mu -ig_2W^\mu_a\tau_a$ and $L=(t_L,b_L)$, $R=(t_R,b_R)$.
Due to their smallness we have ignored all tiny Yukawa couplings and we
will even drop the bottom Yukawa coupling from now on. It is easily verified
that $\delta{\cal L}_{custodial}$ does not spoil $\rho_{tree}\equiv 1$
upon symmetry breaking (i.e. $\Omega=v~{\bf 1\!\!\!1}+\delta\Omega$) even
though the $SU(2)_R$ symmetry is destroyed. Consequently custodial $SU(2)$
violating vertices enter only via loops into the renormalization of the
Higgs sector and guarantee an expansion of the form
\beq
\rho = 1 + G(g_1^2,g_t^2,...) = 1 + c_1 g_1^2 + c_t g_t^2 ~,
\label{rhoexpand}
\eeq
where $G$ is a homogeneous function in $g_1^2$ and $g_t^2$. Note that $c_1$
and $c_t$ depend in general on all other couplings, but eq.~(\ref{rhoexpand})
guarantees in a perturbative expansion Veltman screening, i.e. the
dependence on $\lambda$ (i.e. $m_H^2$) is reduced compared to naive
expectations \cite{Veltscreen}.

The coefficient $c_t$ arises in the \SM at the one loop level from the
diagrams shown in Fig.~\ref{F1}. Numerically\footnote{Note that $c_1$
and $c_t$ are defined without powers of coupling constants.} $c_t$ is
typically about four times bigger than $c_1$ and the current direct
lower limit on the top quark mass \cite{CDF} of $m_t>91~GeV\simeq m_Z$
leads to $g_1^2/g_t^2 = m_Z^2/m_t^2~2\sin^2\theta_W < 2\sin^2\theta_W
\simeq 0.46$, i.e. $g_t^2\gta 2.2~g_1^2$. The biggest correction to
$\rho=1$ comes therefore from the top quark and the experimental data
for $\Delta\rho^{exp}$ can be translated into a prediction of $g_t$,
i.e. the top mass:
\beq
\Delta\rho^{exp}= \Delta\rho^{theo}
\simeq \frac{N_c}{32\pi^2}~g_t^2=\frac{N_c}{32\pi^2}~\frac{m_t^2}{v^2}
=\frac{N_c\alpha_{em}m_t^2}{16\pi\sin^2\theta_W\cos^2\theta_WM_Z^2}~.
\label{predict}
\eeq
This is actually the dominating effect in top mass predictions based on the
analysis of radiative corrections of the \SMp This leads today for
$m_H=300~GeV$ to
$m_t = 152 {\mbox{\scriptsize$
\begin{array}{c} +18 \\ -20 \end{array}
\left(
\begin{array}{cccccc}
+17 & {\rm for} & m_H & = & 1  & TeV \\
-21 & {\rm for} & m_H & = & 60 & GeV
\end{array}
\right)$}}~GeV$~\cite{radiative}.

It is possible that the top quark is not precisely found where required
by the \SM and therefore corrections to $\Delta\rho$ from new physics
should be studied. We discuss here modifications of custodial $SU(2)$
violation due to a possible dynamical origin of fermion masses. If e.g.
the top mass has dynamical origin then $m_t$ is replaced by a dynamical
top mass function $\Sigma_t(p^2)$ while the physical top mass is given
by one point only, namely the solution of the on--shell--condition
$m_t=\Sigma(m_t^2)$. In Section II we calculate $\Delta\rho$ for an
arbitrary fermionic weak isospin doublet with momentum dependent mass
functions $\Sigma_i(p^2)$. In Section III we present some limiting cases
and illustrate magnitude and sign of typical modifications. We show that
relative to the \SM positive and negative corrections to $\Delta\rho$
can occur and we will point out that it is in principle possible to keep
$\Delta\rho$ fixed while the physical top mass can essentially take any
value. In Section IV we relate these results to the type of gap equation
and show that this may provide in a certain class of models a natural
compensation mechanism which makes $\Delta\rho$ systematically smaller
than expected. The implications for \EW precision measurements on general
\DSB scenarios are discussed in Section V.

\section{$\Delta\rho$ for Dynamical Fermion Masses}

Suppose the Higgs sector is replaced by some dynamical scenario which is
responsible for the breaking of the \EW symmetry and for quark and lepton
masses. Consequently the underlying Lagrangian would be the \SM without
the Higgs sector\footnote{I.e. just kinetic terms for quarks, leptons and
$U(1)_Y\times SU(2)_L\times SU(3)_C$ gauge fields.} amended by a new
(presumably strongly coupled) sector triggering dynamical symmetry breaking.
This new sector may contain new fundamental fermions and/or bosons, but
may also stand for an effective description of non--perturbative effects
of the known fermions and gauge fields. In any case there must be a scalar
operator which develops a condensate (or VEV) such that the broken global
symmetries give rise to those \GBs which can give mass to $W$ and $Z$. Well
known examples are Technicolor \cite{TC,ETC}, top condensation \cite{topC}
and even the \SM Higgs mechanism can be phrased in this way.

Besides breaking the $SU(2)_L$ gauge symmetry, \DSB (DSB) should also
explain fermion masses like those which arise via Yukawa interactions in
the \SMp When fundamental scalars are absent this is achieved by connecting
the fermions in a suitable way to some \EW symmetry breaking fermionic
condensate. A given fermion is therefore either condensing itself such
that its mass is the result of a ``critical'' Schwinger--Dyson (or gap)
equation or alternatively the fermion is coupled indirectly via some (e.g.
``see--saw'' or ``horizontal'') interaction to the condensation mechanism.
In both cases the fermion masses become therefore momentum dependent
functions $\Sigma(p^2)$ related directly or indirectly to some gap equation.
For an asymptotically free condensing force $\Sigma(p^2)$ approaches
zero at high momenta $p^2\rightarrow\infty$. This asymptotic behaviour
starts typically around some generic DSB scale, which -- if the underlying
condensation mechanism is to solve the old hierarchy problem -- should
not be many $TeV$. Note that this should imply structure in $\Sigma(p^2)$
at a few $TeV$.

We assume now that symmetry breaking is the result of unspecified new
strong forces acting on some fermion doublet(s) and that -- like in
the \SM -- custodial $SU(2)$ violation does not change significantly
if the weak $U(1)_Y$ coupling $g_1$ is set to zero\footnote{Indirectly
(via vacuum alignment) a small custodial $SU(2)$ violating $U(1)_Y$
coupling can however become very important.}. In that limit custodial
$SU(2)$ breaking must stem entirely from the new sector which is coupled
to the $W_3$ and $W_\pm$ propagators only via those fermions which are
representations under both $SU(2)_L$ and the new strong force. All
custodial  $SU(2)$ violations arise then from the contributions of
fermionic vacuum polarizations to the $W$ propagator.
In an expansion in powers of $g_2^2$ the leading contribution is
given by fermion loop corrections to the $W$ propagator which do not
contain any \EW gauge boson propagation inside the loop. Insertions of
fermionic vacuum polarizations into \EW loop diagrams are suppressed
by corresponding powers of $g_2^2$. In leading order $g_2^2$, but exact
in the new strong coupling, the custodial $SU(2)$ violating contributions
to the $W$ propagator are graphically represented in Fig.~\ref{F2}.
The first contribution is the generalization of the type of diagram
shown in Fig.~\ref{F1} with hard masses replaced by $\Sigma$'s, i.e. all
diagrams which contribute to the dynamically generated fermion masses.
The second contribution contains the exact Kernel $K$ of the strong forces
responsible for condensation and it is useless to expand this Kernel
perturbatively in powers of the coupling constants of the new strong force.
The Goldstone theorem tells us however that the Kernel must contain poles
of massless \GBs due to the global symmetries broken by the fermionic
condensates. This is symbolically expressed by the second line of
Fig.~\ref{F2}, where $\tilde K$ does not contain any further poles of
massless particles. But $\tilde K$ may (and typically will) contain
all sort of massive bound states like vectors, Higgs--like scalars etc.
in all possible channels.

The \GB contributions\footnote{Which are very important for a gauge
invariant dynamical Higgs mechanism.} shown in Fig.~\ref{F2} were used by
Pagels and Stokar \cite{PaSto} to obtain a relation between the $\Sigma$'s
and the \GB decay constant. Their derivation uses the fact that only the
\GBs contribute a term proportional $p_\mu p_\nu/p^2$ to the $W$
polarization at vanishing external momentum, but this method ignores possible
contributions from $\tilde K$ which enter indirectly via the use of Ward
identities. The $p_\mu p_\nu/p^2$ contributions to $\Pi_{\mu\nu}$ are
balanced (up to small corrections from $\tilde K$) by $g_{\mu\nu}$ terms
created by the first diagram on the {\em rhs} of Fig.~\ref{F2}. We derive
now a relation between the $\Sigma$'s and the \GB decay constants from
these $g_{\mu\nu}$ terms and compare the result later with the Pagels Stoker
relation. We will further argue that contributions from $\tilde K$
are significantly suppressed. Let us therefore work with rescaled fields
such that gauge couplings appear in the kinetic terms of the gauge boson
Lagrangian like $\left(-1/4g^2\right)\left(W_{\mu\nu}\right)^2$.
Since we do not include any propagating $W$ bosons we need not gauge fix
at this stage and the inverse $W$ propagator can be written as
\beq
\frac{1}{g_2^2}D_{W,\mu\nu}^{-1}(p^2) = \frac{1}{g_2^2}
  \left(-g_{\mu\nu}+\frac{p_\mu p_\nu}{p^2}\right)~p^2
- \Pi_{\mu\nu}(p^2)~,
\label{invprop}
\eeq
with the polarization tensor
$\Pi_{\mu\nu}(p^2)=\left(-g_{\mu\nu}p^2+p_\mu p_\nu\right)\Pi(p^2)$.
At vanishing external momentum the first fermion loop on the {\em rhs}
of Fig.~\ref{F2} contributes to $\Pi_{\mu\nu}$
\beq
\Pi_{\mu\nu} = -iZ^2N_c~\int\frac{d^4k}{(2\pi)^4}~
\frac{
{\sl Tr}\left[\Gamma_\mu(\slash k+\Sigma_1(k))\Gamma_\nu
(\slash k+\Sigma_2(k))\right]
}
{(k^2-\Sigma_1(k)^2)(k^2-\Sigma_2(k)^2)}~,
\label{Pimunu}
\eeq
where $Z^{-1}=\sqrt{2}, 2$ in the charged and neutral channel, respectively,
$\Gamma_\alpha=(1-\gamma_5)\gamma_\alpha$, and ${\bf +i\epsilon}$ is
generally implied in the denominator. By naive power counting
eq.~(\ref{Pimunu}) has quadratic and logarithmic divergences, but
assuming\footnote{This is justified for asymptotically free theories
where chiral symmetry breaking disappears as $p^2\rightarrow\infty$.}
$\Sigma_i(p^2)\stackrel{p^2\rightarrow\infty}{\bf\longrightarrow}0$ we
find that the divergences of $\Pi_{\mu\nu}(p^2)$ are identical to those
calculated for $\Sigma_i\equiv0$. It makes therefore sense to split
$\Pi_{\mu\nu}(p^2)=\Pi^0_{\mu\nu}(p^2)+\Delta\Pi_{\mu\nu}(p^2)$ where
$\Pi^0_{\mu\nu}$ is defined as $\Pi_{\mu\nu}$ for $\Sigma_i\equiv0$.
$\Pi^0_{\mu\nu}$ is then an uninteresting $\Sigma_i$ independent constant
which contains all divergences and needs renormalization. Contrary the
interesting $\Sigma_i$ dependent piece
$\Delta\Pi_{\mu\nu}=\Pi_{\mu\nu}-\Pi^0_{\mu\nu}$
is finite, even when the external momentum is sent to zero. Thus
\bea
\Delta\Pi_{\mu\nu} &=&
  -iZ^2N_c~\int\frac{d^4k}{(2\pi )^4}~\left\{
    \frac{
    {\sl Tr}
    \left[\Gamma_\mu(\slash{k}+\Sigma_1)\Gamma_\nu(\slash{k}+\Sigma_2)\right]
           }{(k^2-\Sigma_1^2)(k^2-\Sigma_2^2)}
  -
    \frac{
    {\sl Tr}\left[\Gamma_\mu\slash{k}\Gamma_\nu\slash{k}\right]
           }{k^4}
                                 \right\} \, ,\\
                  &=&
  -iZ^2N_c~\int{}\frac{d^4k}{(2\pi)^4}~
  {\sl Tr}\left[\Gamma_\mu\slash{k}\Gamma_\nu\slash{k}\right]
  \left\{
  \frac{1}{(k^2-\Sigma_1^2)(k^2-\Sigma_2^2)}-\frac{1}{k^4}
  \right\} \nonumber \\
                 & &
  -iZ^2N_c~\int\frac{d^4k}{(2\pi)^4}~
  \Sigma_1\Sigma_2 {\sl Tr}\left[\Gamma_\mu\Gamma_\nu\right]
  \left\{
  \frac{1}{(k^2-\Sigma_1^2)(k^2-\Sigma_2^2)}
  \right\}~,
\label{DeltaPi}
\eea
where $N_c$ is the number of colors and $\Gamma_i=(1-\gamma_5)\gamma_i$.
Note that our separation procedure for $\Delta\Pi_{\mu\nu}$ will not
spoil gauge invariance. The first trace in eq.~(\ref{DeltaPi}) gives
under the integral $-\frac{1}{2}g_{\mu\nu}k^2$ while the second trace
vanishes. Angular integration is trivially performed in Euclidean space
and continued back to Minkowski space:
\beq
\Delta\Pi_{\mu\nu} = - g_{\mu\nu}~\frac{Z^2N_c}{(4\pi)^2}
\int\limits_0^\infty dk^2~
\frac{k^2(\Sigma_1^2+\Sigma_2^2) - \Sigma_1^2\Sigma_2^2}
{(k^2-\Sigma_1^2)(k^2-\Sigma_2^2)}~.
\label{finalDPi}
\eeq
As anticipated this result is homogenous in $\Sigma_i$ and finite with the
assumptions made on $\Sigma_i$. For neutral channels eq.~(\ref{finalDPi})
must be summed over all fermion anti--fermion pairs with $\Sigma_1=\Sigma_2$
and for charged channels one must sum over all doublets, where $\Sigma_1$
and $\Sigma_2$ represent then the fermion masses of the isospin doublet.
We can for example neglect the bottom quark mass for the contribution of
the $t-b$ doublet and set $\Sigma_1=\Sigma_2=\Sigma_t$ in the neutral
channel and $\Sigma_1=\Sigma_t$, $\Sigma_2=\Sigma_b\equiv 0$ in the charged
channel, respectively. The contributions of any other fermion doublet are
given by the same formula provided $N_c$ is suitably replaced.

The \GB decay constants $F_i^2$ are the poles of $\Pi(p^2)$ at vanishing
external momentum. For our definition of $\Pi_{\mu\nu}$ we find that
$F_i^2$ is identical to the $g_{\mu\nu}$ piece eq.~(\ref{finalDPi})
without the factor $-g_{\mu\nu}$. Taking into account $Z=1/\sqrt{2}$ in
the charged channel and $Z=1/2$ in the neutral channel and allowing for
further arbitrary custodial SU(2) symmetric contributions $F_o^2$ one finds
\bea
F_\pm^2 &=& F_0^2 + \frac{N_c}{32\pi^2}\int\limits_0^\infty dk^2~
            \frac{k^2(\Sigma_1^2+\Sigma_2^2)-\Sigma_1^2\Sigma_2^2}
            {(k^2-\Sigma_1^2)(k^2-\Sigma_2^2)}
                                                               \nonumber \\
&\stackrel{\slimits}{\longrightarrow}&
            F_0^2 + \frac{N_c}{32\pi^2}\int\limits_0^\infty dk^2~
            \frac{\Sigma_t^2}{k^2-\Sigma_t^2}~,
                                                             \label{Fpm} \\
F_3^2   &=& F_0^2 + \frac{N_c}{32\pi^2}\int\limits_0^\infty dk^2~\left\{
            \frac{k^2\Sigma_1^2-\frac{1}{2}\Sigma_1^4}{(k^2-\Sigma_1^2)^2}+
            \frac{k^2\Sigma_2^2-\frac{1}{2}\Sigma_2^4}{(k^2-\Sigma_2^2)^2}
                                                          \right\}
                                                               \nonumber \\
&\stackrel{\slimits}{\longrightarrow}&
            F_0^2 + \frac{N_c}{32\pi^2}\int\limits_0^\infty dk^2~
            \frac{k^2\Sigma_t^2-\frac{1}{2}\Sigma_t^4}{(k^2-\Sigma_t^2)^2}~,
                        \label{FF3}
\eea
such that
\beq
F_3^2-F_\pm^2 =
\frac{N_c}{64\pi^2}\int\limits_0^\infty dk^2
\frac{k^4(\Sigma_1^2-\Sigma_2^2)^2}{(k^2-\Sigma_1^2)^2(k^2-\Sigma_2^2)^2}
%                                                   \label{dF2}\nonumber \\
\stackrel{\slimits}{\longrightarrow}
\frac{N_c}{64\pi^2}
\int\limits_0^\infty dk^2~\frac{\Sigma_t^4}{(k^2-\Sigma_t^2)^2}~.
                                                        \label{finaldF2}
\eeq
Eq.~(\ref{Fpm}) for $F_\pm^2$ is equivalent to the result obtained by Pagels
and Stokar \cite{PaSto} from the $q_\mu q_\nu/q^2$ contributions of \GBs
to $\Pi_{\mu\nu}$. The result for the neutral channel, eq.~(\ref{FF3}),
looks however somewhat different. By using the integral identity
\beq
\int\limits_0^\infty dx ~\frac{x^2f(x)^\prime - f(x)^2}
{\left( x-f(x) \right)^2} = f(\infty )~,
\label{intid}
\eeq
for $x=k^2$ and $f=\Sigma_i^2$ we can rewrite eq.~(\ref{FF3}) for example
in the case $\Sigma_1=\Sigma_t$, $\Sigma_2=0$
\beq
F_3^2 = F_0^2 + \frac{N_c}{32\pi^2}\int\limits_0^\infty dk^2~k^2~
      \frac{\Sigma_t^2- k^2\Sigma_t\Sigma_t^\prime}{(k^2-\Sigma_t^2)^2}~,
\label{PASTOres}
\eeq
where $\Sigma_t^\prime = d\Sigma_t /dk^2$. Even though this looks now
formally similar to the Pagels Stokar result it differs by a factor 2
in front of the derivative term in the nominator of eq.~(\ref{PASTOres}).
This difference may appear less important, but we will see in Section III
that in the limit of a hard top mass our method produces the correct
$\rho$--parameter, while the Pagels Stokar result produces 3/2 times the
correct answer. In addition to the correct $\rho$--parameter limit our
expression leads also to a better numerical estimate of $f_\pi$ if we
follow the methods of ref.~\cite{PaSto}. The difference between our result
and the Pagels Stokar result must be resolved by $g_{\mu\nu}$ and
$q_\mu q_\nu/q^2$ contributions to $\Pi_{\mu\nu}$ from $\tilde K$ in the
second line of Fig.~\ref{F2} such that the full result is transverse.

The $\rho$--parameter can be rewritten as
\beq
\rho = 1+\Delta\rho =
\frac{F_\pm^2}{F_3} = \left(1+\frac{(F_3^2-F_\pm^2)}{F_\pm^2}\right)^{-1}
\simeq 1 - 2~\frac{(F_3^2-F_\pm^2)}{v^2}~,
\label{rhoaprox}
\eeq
and from eq.~(\ref{finaldF2}) we find the contribution of any fermion
doublet\footnote{I.e. this formula applies to many cases such as
for example for Technicolor.} to the $\rho$--parameter
\beq
\Delta\rho=\frac{-N_c}{32\pi^2v^2}~\int\limits_0^\infty dk^2~
   \frac{k^4(\Sigma_1^2-\Sigma_2^2)^2}{(k^2-\Sigma_1^2)^2(k^2-\Sigma_2^2)^2}
\stackrel{\slimits}{\longrightarrow}
\frac{-N_c}{32\pi^2v^2}~\int\limits_0^\infty dk^2~
   \frac{\Sigma_t^4}{(k^2-\Sigma_t^2)^2}~,
\label{finalrho}
\eeq
where we used $F_\pm^2=v^2/2$ with $v\simeq 175~GeV$ in the denominator
and the custodial $SU(2)$ symmetric contributions $F_0^2$ have disappeared
as they should.

The final expression for the $\Delta\rho$ implies that for given
$\Sigma_t(p^2)\stackrel{p^2\rightarrow \infty}{\longrightarrow}0$ we can
calculate three observable quantities which are one of the \GB decay
constants $F_i^2$, $\Delta\rho$ and furthermore the physical top mass
defined via $\Sigma_t(m_t^2)=m_t$. These three quantities are dominated
by different momenta and therefore $\Sigma\neq constant$ leads to a
different answer than a constant, i.e. hard mass. In this context
it is instructive to look at the degree of convergence of the above
integrals. The \GB decay constants\footnote{They are related to the $W$--
and $Z$-- masses via $M_W^2=g_2^2F_\pm^2$ and $M_Z^2=g_2^2F_3^2 =
(g_1^2+g_2^2)F_Z^2$} $F_i^2$ are formally log. divergent, but are finite
with our assumption on $\Sigma_t(p^2)$. In that case renormalization is
not needed, but due to the formal log. divergence $\Sigma$ contributes
with equal weight at all momentum scales. In other words, the magnitude
of $F_i^2$ depends crucially on the high energy tail of $\Sigma_i$. The
difference $F_\pm^2 - F_0^2$ has better convergence properties and is
always finite, even for $\Sigma_t(p^2)=constant$. This implies that
$\rho$ is finite, as it should be, and it is most sensitive to infrared
scales somewhat above $m_t$. We will illustrate now effects of structure
in $\Sigma$ and postpone a discussion how certain $\Sigma$ emerge from a
gap equation of the underlying dynamics in Section IV.

\section{Magnitude and Sign of Effects}

The result eq.~(\ref{finalrho}) for $\Delta\rho$ has several interesting
limiting cases. First we would like to see if the correct \SM result
emerges for a $t-b$ doublet. Therefore we set
\beq
\Sigma_t(p^2) = m_t\,\Theta(\Lambda^2-p^2)~,
\label{topan1}
\eeq
and ignore the $b$ quark mass. From eq.~(\ref{finalrho}) we obtain
\beq
\Delta\rho
= -\frac{N_c}{32\pi^2v^2}~\int\limits_0^{\Lambda^2}
                               dk^2~\frac{m_t^4}{(k^2-m_t^2)^2}
= \frac{N_cm_t^2}{32\pi^2v^2}\left(\frac{1}{1-m_t^2/\Lambda^2}\right)~.
\label{rho1}
\eeq
which becomes in the limit $\Lambda\rightarrow\infty$ (i.e. a hard, constant
top mass)
\beq
\Delta\rho
= \frac{N_c}{32\pi^2}~\frac{m_t^2}{v^2}
= \frac{N_c\alpha_{em}}{16\pi\sin^2\theta_W\cos^2\theta_W}
  ~\frac{m_t^2}{M_Z^2}~,
\label{finalrho1}
\eeq
which is correctly the leading \SM result. Note that the Pagels Stokar
relation produces in this limit incorrectly 3/2 times the \SM result
while our expression gives the correct answer. For finite $\Lambda$
eq.~(\ref{rho1}) describes furthermore the modification of the \SM result
due to a high energy momentum cutoff
\beq
\Delta\rho_\Lambda = \Delta\rho_{SM}
\left(\frac{1}{1-m_t^2/\Lambda^2}\right)
\simeq
\Delta\rho_{SM}
\left( 1+m_t^2/\Lambda^2 \right)~,
\label{rhocut}
\eeq
where the last simplification is valid for $m_t\gg\Lambda$. The
cutoff\footnote{Which may not only stand for the falloff of $\Sigma$ but
also for some other cancellation mechanism.} makes $\Delta\rho$ more positive
than in the \SM which implies for a fixed experimental value of $\Delta\rho$
a lower top mass prediction. An ansatz like eq.~(\ref{topan1}) can be
viewed as the result of the Nambu--Jona-Lasinio gap equation of top
condensation \cite{BHL} and exhibits the leading correction to
$\Delta\rho_{SM}$ for such models.

The corrections to $\Delta\rho$ can in principle also go into the opposite
direction. Consider for example a modification of the above ansatz
\beq
\Sigma_t(p^2) = \left\{
\begin{array}{ll}
m_t               & {\rm for~} p^2 < \Lambda_1^2;~ \Lambda_1^2 > m_t^2;\\
\sqrt{r}\cdot m_t & {\rm for~} \Lambda_1^2 \leq p^2 \leq \Lambda^2;\\
0                 & {\rm for~} p^2 > \Lambda^2;
\end{array}\right.
\label{topan2}
\eeq
where $\Sigma$ is enhanced $r$--fold above $\Lambda_1<\Lambda$ before it
vanishes at $\Lambda$ as before. The modified result is
\bea
\!\!\!\!\!\!
\Delta\rho &=& \frac{N_cm_t^2}{32\pi^2v^2}
\left(\frac{1}{1-m_t^2/\Lambda^2}
   -\left[
      \frac{r^2m_t^2(\Lambda^2-\Lambda_1^2)}
                {(\Lambda^2-rm_t^2)(\Lambda_1^2-rm_t^2)}
      -\frac{m_t^2(\Lambda^2-\Lambda_1^2)}
                {(\Lambda^2-m_t^2)(\Lambda_1^2-m_t^2)}
   \right] \right)~,~ \label{rho2} \\
& & \stackrel{\Lambda^2,\Lambda_1^2\gg m_t^2, rm_t^2}{\bf\simeq}
    \frac{N_cm_t^2}{32\pi^2v^2}
    \left(1+m_t^2/\Lambda^2
        -\left[\frac{m_t^2(\Lambda^2-\Lambda_1^2)}
                    {\Lambda^2\Lambda_1^2}~(r^2-1)\right]
    \right) ,      \label{finalrho2}
\eea
where extra contributions due to $r\neq 1$ and $\Lambda_1\neq\Lambda$
are isolated in square brackets.
Compared to eq.~(\ref{topan1}) the ansatz eq.~(\ref{topan2}) has for $r>1$
an extra ``bump'' between $\Lambda_1$ and $\Lambda$. This bump counteracts
the effect of the cutoff and makes $\Delta\rho$ less positive and it
is easy to see that the bump can even become more important than the cutoff.
This illustrates that scales somewhat above $m_t$ are very important
for the magnitude and sign of $\Delta\rho$ and it is natural to ask if
$\Sigma$ can be chosen such that $\Delta\rho$ vanishes for an arbitrary
value of $m_t$. This can indeed be done by choosing for example by hand
\beq
\Sigma_t(p^2) = \frac{241 \left( 118\,m_t^4 + p^4 \right)}
                      {7 \left( 4096\,m_t^6+p^6 \right)}\,m_t^3~,
\label{rho0ex}
\eeq
which is shown graphically in Fig.~\ref{F3} to have only very moderate
structure.

At this point it is necessary to say a few words on the integration over
the pole of eq.~(\ref{finalrho}). Instead of performing an analytic
continuation for any ansatz individually one can rewrite
eq.~(\ref{finalrho}) exactly into
\bea
\Delta\rho
&=& \frac{N_c m_t^2}{32\pi^2v^2}\Bigg( 1+
    \frac{4m_t\Sigma^\prime_t(m_t^2)-4m_t^2\Sigma^\prime_t(m_t^2)^2}
    {\left( 1-2m_t\Sigma^\prime_t(m_t^2)\right)^2}             \nonumber \\
&&  -\int\limits_0^\infty \frac{dk^2}{m_t^2}\Bigg[
    \frac{\Sigma_t(k^2)^4}{\left( k^2-\Sigma_t(k^2)^2\right)^2}
    -\frac{m_t^4}{\left( k^2-m_t^2\right)^2} \cdot
    \frac{1}{\left( 1-2m_t\Sigma^\prime_t(m_t^2)\right)^2}
                                                       \label{approxrho} \\
&&  -\frac{4m_t^5}{k^4-m_t^4} \Bigg(
    \frac{2\Sigma^\prime_t(m_t^2)}
    {\left( 1-2m_t\Sigma^\prime_t(m_t^2)\right)^2}
    +\frac{m_t\Sigma^\prime_t(m_t^2)^2+m_t^2\Sigma^{\prime\prime}_t(m_t^2)}
    {\left( 1-2m_t\Sigma^\prime_t(m_t^2)\right)^3}
    \Bigg)\Bigg]\Bigg)~,                                        \nonumber
\eea
which has the advantage that the integrand in square brackets does not
have an explicit pole for any arbitrary given $\Sigma_t(p^2)$.

In order to illustrate that our result is not just limited to the
contributions of a $t-b$ doublet we can look for example at Technicolor
\cite{TC} where an extra doublet of Techni--fermions $U-D$ condenses and
breaks the \EW symmetry. Ordinary quark and lepton masses (like the
top mass) must be generated by so--called Extended Technicolor \cite{ETC}
interactions\footnote{Which must be settled at very high scales in order
to be compatible with experimental limits on Flavour Changing Neutral
Currents (FCNC).}. The coupled system of gap equations leads in a rough
approximation \cite{AppWi} to the relation $\Sigma_U-\Sigma_D = \Sigma_t$.
Assuming this relation and $\Sigma_i=m_i\,\Theta(\Lambda^2-p^2)$ we obtain
from eq.~(\ref{finalrho})
\beq
\Delta\rho = \frac{N_cm_t^2}{32\pi^2v^2}
\left(1 + \frac{4}{9} N_{TC} + \frac{m_t^2}{\Lambda^2}
+\frac{4}{3} N_{TC}\frac{m_U^2}{\Lambda^2}
+{\cal O}(m^4/\Lambda^4) \right)~,
\label{rhoTC}
\eeq
which becomes for $\Lambda\rightarrow\infty$ the result which is
quoted in the literature \cite{AppWi}. For finite $\Lambda$ we find the
$m_t^2/\Lambda^2$ correction of eq.~(\ref{rhocut}) and additionally a
term proportional to $m_U^2/\Lambda^2$. These $1/\Lambda^2$ terms are
small and are usually omitted. This Technicolor example illustrates that
our result works generally for cases where our assumptions are fulfilled.
$\Delta\rho$ is given as soon as all fermionic doublets, their color
factors and their $\Sigma^\prime$s are known. One might think that this
does not contain much information without specifying a detailed theory,
but we will see that there are interesting model independent consequences.

\section{Reduced $\Delta\rho$ and the Type of Gap Equation}

In the discussion of the previous Section we showed that a $\Sigma_t$ with
a ``bump'' leads to a $\Delta\rho$ which is considerably smaller than
expected from the pole mass. For a fixed experimental value of
$\Delta\rho^{exp}$ this would imply systematically a higher top mass
prediction from radiative corrections if such a bump arises naturally. Such
a bump can be phrased as a negative contribution to $\Delta\rho$
compared to the \SM and since there are only a few known ways to get
such a negative contribution to $\Delta\rho$ we would like to show what
kind of gap equation could lead to such a scenario. We consider therefore
a situation where $\Sigma_t$ arises from the exchange of a boson with mass
$M_X$ as indicated in Fig.~\ref{F4}. The full gap equation Fig.~\ref{F4} is
too complicated and therefore one uses the so--called ladder approximation
Fig.~\ref{F5} which can be written as
\beq
S^{-1}(p)-\slash{p}=\int \frac{d^4k}{(2\pi)^4}\Gamma^a S^{-1}(k)\Gamma_a
   \frac{-i}{(p-k)^2-M_X^2}~,
\label{ladder}
\eeq
where
\beq
S(p)= \frac{i}{\slash{p}-\Sigma (p^2)}~,
\label{SF}
\eeq
is the fermion propagator and $\Gamma^a$ the vertex. The index $a$
runs over all Minkowski and internal group indices corresponding to the
interaction structure. Angular integration leads to
\beq
\Sigma(p^2)=C \int\limits_0^\infty dk^2 \frac{\Sigma(k^2) k^2}
{k^2-\Sigma(k^2)^2} \; K(p^2,k^2,M_X^2)
\;\;\; , \;\;\;\;\;\; C=-\frac{\Gamma^a\Gamma_a}{(4\pi)^2}~,
\label{radialgap}
\eeq
with the Kernel
\beq
K(p^2,k^2,M_X^2)=  \frac{2}{(k^2+p^2-M_X^2)\left(1+\sqrt{1-\frac{k^2 p^2}
                         {(k^2+p^2-M_X^2)^2} }~\right)}~,
\label{Kernel}
\eeq
where $C$ is a constant which depends only on the strength and group
structure of the new interaction.

\noindent
In this approximation exist a number of simple arguments why $\Sigma$
should have a bump when $M_X\neq 0$:
\begin{enumerate}
\item The self--energy graph of the ladder approximation has a resonance
like structure at $p^2=(M_X+m_t)^2$ due to the generation of real particles
above that scale. This explains also why there is no bump in
QCD for momenta higher than the constituent quark masses.
\item Because of this resonance structure in the complex plane there is a
cut on the real axis for momenta $p^2$ higher than $(M_X+m_t)^2$. We demand
that $\Sigma$ is analytic at all other points, which is plausible in ladder
approximation. If $\Sigma$ does not have zeros in the complex plane, we
know from the theory of analytic functions that the maximum of
$\left|\Sigma\right|$ must be at the boundary. Therefore there must be a bump
at the cut since $\left|\Sigma\right|\to 0$ for $\left|p^2\right|\to\infty$.
\item Demanding maximal analyticity one can also use the gap equation in
Euclidean space
\beq
\Sigma(-p^2)=C \int\limits_0^\infty dk^2 \frac{\Sigma(-k^2) k^2}
{k^2+\Sigma(-k^2)^2} \; K(-p^2,-k^2,M_X^2)~,
\label{maxanalytic}
\eeq
and one finds
\beq
\Sigma^\prime(0)=C \int\limits_0^\infty dk^2 \frac{\Sigma(-k^2) k^2}
{k^2+\Sigma(-k^2)^2} ~ \frac{3k^2+4M_X^2}{4(k^2+M_X^2)^3}~,
\label{analyticsigmap}
\eeq
which is positive, even if $\Sigma$ has zeros at large $k^2$.
Furthermore $\Sigma^\prime$ is positive for small $p^2$ which shows also
that there must be a bump.
\end{enumerate}

In solving the gap equation numerically one runs easily into problems
because of the slow decrease of the $\Sigma$-function(s). The integral
equation is best transformed into a discrete eigenvalue problem by dividing
the $k^2$-axis up to a cutoff into $n$ intervals or by using a special
discrete function space, e.g. a Taylor expansion on the M\"obius-transformed
$k^2$-axis. With this methods we found a critical value for $C$ which is
in good agreement with the bound $C_{\rm crit}\,>\,1/4$ derived by
T.~Maskawa and H.~Nakajima \cite{MaNa}. The effects on $\Delta\rho$ are
for reasonable parameters typically $10$-$20$\% corrections to the
\SM value and become biggest when $m_t$ is of the same magnitude as $M_X$.

Clearly such a calculation is not exact but gives only a qualitative
impression of the magnitude of the effects. In principle one can also
calculate the \GB decay constants and the $W$ mass directly. This leads
typically to a result which is to small by a factor $2$. But this can
easily be due to the uncertainty in the asymptotic high energy behaviour
of the solutions of such gap equations. In contrast $\Delta\rho$ does not
get big contributions from the asymptotic part because of the strong
convergence of the integral in eq.~(\ref{finalrho}). Therefore $\rho$ is
not sensitive to the ultra high energy details of $\Sigma$.

The ladder approximation omits a lot of graphs which could in principle be
relevant in the exact gap equation. Important effects could arise
for example for the following reasons:
\begin{enumerate}
\item The analyticity properties are not obvious such that $\Delta\rho$
might even be negative.
\item The feedback of a composite Higgs resonance is ignored in this ladder
approximation which could even be dominating the gap equation if the top
mass (i.e. the Yukawa coupling) is very big. Due to this feedback there
could be a bump at $M_H\approx 2m_t$ allowing a drastically smaller value
of $\Delta\rho$ and therefore a rather high top mass. Such effects could
be relevant in a realization of Nambu's bootstrap idea of \EW symmetry
breaking. In this case there is further amplification since $g_t$ at
the condensation scale is considerably higher than the on--shell value
$g_t(m_t)$. The top quark might therefore condense for a top quark mass
which is even 1.5 to 2 times smaller than naive values.
\end{enumerate}
Despite of all the technical uncertainties we believe that a massive strongly
coupled gap equation should lead to a ``bump'' scenario which might e.g.
play a role in proposed gauge models of top condensation where a strongly
interacting broken gauge group triggers condensation \cite{TOPCOLOR,U1U1}.

\section{Discussion}

We studied effects of dynamical chiral symmetry breaking of the \SM on
custodial $SU(2)$ violation in the limit where the $U(1)_Y$ coupling $g_1$
vanishes and where only fermion doublets contribute. Under the assumption
$\Sigma_i(p^2)\stackrel{p^2\rightarrow\infty}{\bf\longrightarrow}0$
we were able to derive very general results to leading order in $g_2^2$
and -- in principle -- to arbitrary order in the new strong dynamics
by calculating the finite, $\Sigma_i$ dependent $g_{\mu\nu}$ pieces of
the vacuum polarization tensor. However, since we do not know the spectrum
of the theory under consideration we have to restrict ourself to the
leading contribution of a dynamical fermion loop and we ignore possible
$g_{\mu\nu}$ contributions from massive bound states. For a given
scenario one might assume to know the masses $M_j$ and couplings of bound
states and estimate their contributions, but if these states are heavy then
their contributions would typically be suppressed by factors of
$\Sigma_i^2/M_j^2$. Our results, which apply for an arbitrary $SU(2)_L$
doublet of fermions, are formally similar to the old Pagels Stokar
expressions. It turns however out that the difference cannot be explained
by the integral identity eq.~(\ref{intid}) and the difference must find an
explanation in the remaining contributions of $\tilde K$. Since our result
reproduces in the limit of a hard top mass correctly the well known \SM
$\rho$--parameter we believe that it should be better suited for
phenomenological studies.

We emphasized that the three observables $m_t=\Sigma_t(m_t^2)$,
$\Delta\rho$ and one of the \GB decay constants have different
sensitivities to details of $\Sigma_t$. This implies that the
uncertainties which are introduced via truncations made to obtain
approximate solutions of $\Sigma_t$ enter in different ways.
For example in numerical simulations of the problem the asymptotic high
energy tail of $\Sigma_t$ turns out to be very unstable. This implies that
$m_W/m_t$ is very unstable due to the logarithmic sensitivity of this
ratio to the high energy details (for a Technicolor example of this
statement see for example \cite{Frere}). Contrary $\Delta\rho$
is very insensitive to the high energy tail.

We showed that in general it is possible to obtain negative and
positive corrections to $\Delta\rho$ compared to the result of
a hard, constant top mass. Negative contributions to $\Delta\rho$ are
usually hard to obtain and we discussed therefore somewhat the type of gap
equation that could systematically lead to such negative corrections.
This lead to what we called ``bump'' solutions for $\Sigma_t$ which
might be relevant in gauge models of top condensation or some sort
of \EW bootstrap.

We studied custodial $SU(2)$ violations in terms of
$\Delta\rho=\alpha (T-T_0)$ which is less sensitive to model details than
other \EW observables like $S$, $U$ or the $Zb\bar b$ vertex. Note
however, that the $m_t$ dependence of all of these quantities is dominated
by infrared loop momenta. For given $\Sigma_t$ all these observables should
therefore initially be consistent with one single, constant top mass very
close to the pole mass. Only when the precision is increased it may be
possible to measure the contributions of structure in $\Sigma_i$ to
these observables. In this context it should also be mentioned that structure
in
$\Sigma_i$ at some scale $\Lambda$ can also be understood as a synonym
for contributions due to new particle states above the threshold $\Lambda$.

Remarkably there are some completely model independent conclusions.
First we remark that for analytical functions $\Sigma_i$ and the
absence of poles in the first quadrant $\Delta\rho$ can receive only
positive contributions. This can be seen by rewriting eq.~(\ref{finalrho})
in Euclidean space with a positive integrand:
\beq
\Delta\rho=\frac{N_c}{32\pi^2v^2}~\int\limits_0^\infty dk^2~
\left(
\frac{k^2(\Sigma_1^2-\Sigma_2^2)}{(k^2+\Sigma_1^2)(k^2+\Sigma_2^2)}
\right)^2~.
\label{positivrho}
\eeq
This positivity may in principle be arbitrarily weak and does especially
{\em not} forbid that $\Delta\rho$ is smaller than in the \SMp
An example which illustrates this point was given by the ``bump''
solution. In terms of the variable $T$ this implies that fermionic
contributions can only lead to $T>T_0\simeq -0.7$ whatever
the details of the model are.

Next there are further general features of the corrections to $\Delta\rho$
even without a specific theory. These are -- like in the case of the
Technicolor example -- multiplicative corrections to the \SM value
of $\Delta\rho$ which are either counting with appropriate weights
the number of involved fermions and/or terms $m_t^2/\Lambda^2$ which
are sensitive to structure in $\Sigma_i$. These days it is often said
that Technicolor is phenomenologically in trouble due to the $S$ parameter.
We would like to emphasize that the $T$ parameter will soon become much
more important due to the $4/9~N_{TC}$ correction in eq.~(\ref{rhoTC}).
This term which counts extra fermions will essentially be forbidden if
the lower top mass limits increase further. In a more general context this
fermion counting depends of course on the way how the gap equations are
coupled. Typically there are corrections which count the fermionic degrees
of freedom and the weight should not be very tiny. It is however possible
to build models where this counting is completely absent. In that case
there are only $m_t^2/\Lambda^2$ corrections due to structure
in $\Sigma_t$. But even then one can make interesting conclusions
by precision comparisons of theory and experiment.
If the top mass were discovered somewhat outside the actual \SM window
then this could be due to structure in $\Sigma_i$ which is
(besides other possibilities like Higgs triplets, $W^\prime$ and/or
$Z^\prime$ etc.) another important way to bring experimental
and theoretical values of $\Delta\rho$ in agreement. On the other side
the absence of any mismatch can be used to limit contributions of new
physics -- including fermion counting and the scale where structure
could show up. If one assumes for example the absence of fermion counting
and for $\Sigma_t$ the ansatz eq.~(\ref{topan1}) then the mismatch
between the \SM expectation $m_{t,SM}$ from radiative corrections
and the physical top mass $m_t$ translates into a bound for $\Lambda$
\beq
\Lambda^2 \lta \frac{m_{t,phys}^4}{m_{t,SM}^2-m_{t,phys}^2}~.
\label{Lbound}
\eeq
This bound is shown in Fig.~\ref{F6} as dashed line and a comparison
of the theoretical predicted top mass with its experimental value to
$5~GeV$ would imply sensitivity to scales $\Lambda\simeq 0.5~TeV$.
For arbitrary shapes of $\Sigma_t$ such a bound does of course strictly
speaking not exist, but without fine--tuning of the shape one will
always find a similar bound. If we take for example the numerical
solutions of the gap eq.~(\ref{ladder}) and identify $\Lambda\simeq M_X$
then we obtain the even more interesting solid line of Fig.~(\ref{F6}).
But it will be hard to find such a deviation as long as the Higgs and
top mass are not known precise enough. For the future it is however
conceivable that the top mass is known very precisely from the $\bar tt$
threshold, that the Higgs mass is roughly known (or at least stronger
bounded) and that the theoretical precision of radiative corrections
is a small part of a percent. In that case it is possible to come
to $\Delta m_t$ values below $1~GeV$ or even a few hundred $MeV$ which
would probe extremely interesting $\Lambda$ values.

\vskip 1.cm

\noindent
Acknowledgements: We would like to thank H.G.~Dosch, D.~Gromes,
P.~Haberl, N.~Krasnikov, D.~Ross and B.~Stech for useful discussions.

%------------------------------------------------------------------------
%
% References

%------------------------------------------------------------------------

% ===> Figures are appended as UUENCODED Postscript file; just captions

\newpage

\def\listfigurename{Figure Captions}
\listoffigures

\begin{figure}[htb] \vspace{.1cm}
\caption{{\sl The graphs which are responsible for the leading $m_t$
dependence of $\Delta\rho$ in the \SMp}}
\label{F1}

\caption{{\sl The $W$ propagator in leading order $g_2^2$ and exact
in the new non--perturbative interactions. Fermionic self--energies
are represented as fat dots and the four--fermion Kernel $K$
is represented by a fat circle. In the second line the Kernel is
split into \GB contributions (which arise due to the broken global
symmetries with some non--trivial vertex function) and $\tilde K$
(which has no further massless poles).}}
\label{F2}

\caption{{\sl Example for a $\Sigma_t(p^2)$ with $\Delta\rho\equiv 0$. Note
that a rather mild ``bump'' can already result in big corrections.
The on--shell mass is given by the intersection with the dashed line
$\Sigma(p^2)=p$.}}
\label{F3}

\caption{{\sl The massive gap (Schwinger Dyson) equation involves the
exact contribution from the new interactions carried by a heavy $X$
boson (with unspecified quantum numbers). The full one particle
irreducible fermionic self--energy is given in terms of the full
propagators and the vertex function.}}
\label{F4}

\caption{{\sl Ladder approximation of Fig.~\ref{F4}.\hfill\ }}
\label{F5}

\caption{{\sl Limit on $\Lambda$ from precision measurements of $m_t$.
Shown are the scales $\Lambda$ which are tested by a precision comparison
of $m_t$ as predicted from radiative corrections with the physical
top mass. The dashed line represents eq.~(\ref{Lbound}) while the solid
line was obtained from our numerical simulations of eq.~(\ref{ladder})
where we identify $\Lambda=M_X$.}}
\label{F6}

\vfill

\end{figure}

\end{document}